\documentclass[conference]{IEEEtran}

\IEEEoverridecommandlockouts

\usepackage[T1]{fontenc}
\usepackage{graphicx}
\usepackage{textcomp}
\usepackage{xcolor}
\usepackage{multirow}
\usepackage{tabularx}
\usepackage{booktabs}
\usepackage{makecell}
\usepackage{pifont}
\usepackage{array}
\usepackage{algpseudocode}
\usepackage{algorithm}
\usepackage{stfloats}
\usepackage{placeins}
\usepackage{listings}
\usepackage[most]{tcolorbox}
\usepackage{hyperref}

\newcommand{\good}{\textcolor{green}{\ding{51}}}
\newcommand{\bad}{\textcolor{red}{\ding{55}}}
\newcommand{\na}{\textemdash}
\newcommand{\subheading}[1]{\noindent{\textbf{#1}}}

\begin{document}

\title{APT-Agent: Automated Penetration Testing using Large Language Models}

\author{
\IEEEauthorblockN{William Guanting Li}
\IEEEauthorblockA{
University of Queensland\\
Brisbane, Australia\\
guanting.li@uq.edu.au
}
\and
\IEEEauthorblockN{Alsharif Abuadbba}
\IEEEauthorblockA{
CSIRO Data61\\
Sydney, Australia\\
sharif.abuadbba@csiro.au
}
\and
\IEEEauthorblockN{Kristen Moore}
\IEEEauthorblockA{
CSIRO Data61\\
Melbourne, Australia\\
kristen.moore@csiro.au
}
\and
\IEEEauthorblockN{Dan Dongseong Kim}
\IEEEauthorblockA{
University of Queensland\\
Brisbane, Australia\\
dan.kim@uq.edu.au
}
}
\maketitle

\begin{abstract}
Penetration testing is essential to securing modern web infrastructures, yet traditional manual methods struggle to keep pace with their scale and complexity. Large Language Models (LLMs) offer new opportunities for automating these tasks, but existing approaches face two persistent challenges: hallucination of technical entities and insufficient long-term contextual memory. To address these issues, we present APT-Agent, a fully automated LLM-driven penetration testing framework that systematically orchestrates reconnaissance, exploitation, and exfiltration. APT-Agent introduces a hybrid rectification module to recover hallucinated commands and a command-specific memory architecture to preserve operational context across multi-step attack sequences. We evaluate our APT-Agent on Metasploitable 2 against seven vulnerable services spanning web, database, and network protocols. APT-Agent achieves an 84.29\% end-to-end exploitation success rate, compared to 48.57\% (Script Kiddie) and 18.57\% (PentestGPT) under matched conditions. By reducing cognitive burden and minimizing reliance on human intervention, APT-Agent represents a step toward scalable, reliable, and cognitively efficient automation for penetration testing.
\end{abstract}

\begin{IEEEkeywords}
Automated Penetration Testing, LLM, Penetration Testing, Cybersecurity
\end{IEEEkeywords}

\section{Introduction}
In the realm of cybersecurity, penetration testing (pen-testing) plays a vital role by proactively simulating authorized attacks to evaluate and strengthen the resilience of information systems \cite{b1}. However, as digital infrastructures continue to grow in scale and complexity, traditional manual approaches face significant limitations. These include challenges of scalability, dependence on scarce human expertise, and variability in outcomes across practitioners \cite{b18}. Such constraints not only hinder timely and comprehensive assessments but also contribute to cognitive overload among security professionals, diverting their focus from higher-order analysis and strategic defense planning \cite{jiang2024automatedprogressiveredteaming}. Taking these challenges into consideration, automated approaches have the potential to address many of these challenges while preserving accuracy and trustworthiness \cite{AutoPentestingAnOverview}.

\begin{figure}[h]
\centering
\includegraphics[width=1.0\linewidth]{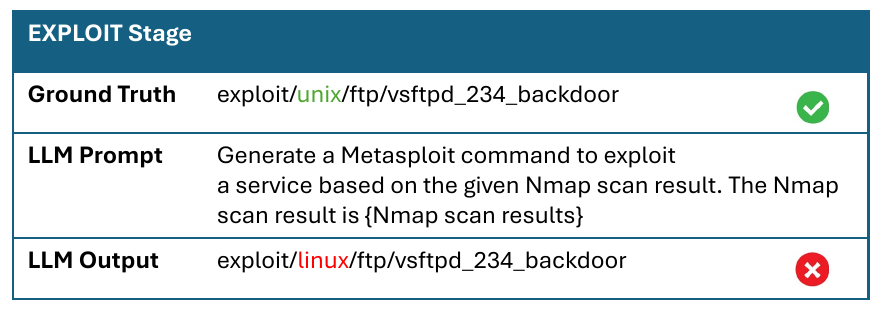}
\caption{Example of Hallucination}
\label{fig:hallucination_example}
\end{figure}

Large Language Models (LLMs) have demonstrated remarkable capabilities in interpreting instructions, drawing inferences, and generating coherent, domain-specific responses \cite{zhao2023survey,Liu_2023}. These strengths extend their applicability well beyond general conversation, positioning them as promising tools for specialized technical domains such as cybersecurity \cite{mayoralvilches2023exploitflowcybersecurityexploitation,zhang2023doesllmgeneratesecurity,he2025largelanguagemodelsblockchain,abuadbba2026promise}. By automating routine and resource-intensive components of pen-testing, LLMs reduce the burden on human operators and allow them to dedicate more time to critical decision-making and complex threat analysis. Recent research has begun to explore the integration of LLMs into pen-testing workflows. Fang et al. \cite{b3,b4} demonstrated LLM agents capable of autonomously discovering and exploiting vulnerabilities in websites and one-day CVEs. Deng et al. \cite{b6} introduced the Pentesting Task Tree (PTT) to structure LLM-driven decisions, while Xu et al. \cite{b7} proposed a Planner, Navigator, Summarizer pipeline enhanced with retrieval-augmented generation (RAG), which also explored whether LLMs can simulate hands-on-keyboard post-breach actions, including lateral movement, credential harvesting, and persistence mechanisms. Despite these advances, current systems still consist of the following two key challenges:

\subheading{Challenge \#1: Hallucination of technical entities}. LLMs frequently invent or misstate particular technical identifiers. For example, exact command syntaxes, Metasploit module paths, module types (exploit/auxiliary/post), target platform names, or payload identifiers \cite{zhang2023languagemodelhallucinationssnowball,li2024drowzeemetamorphictestingfactconflicting,manakul2023selfcheckgpt}. As illustrated in Fig.~\ref{fig:hallucination_example}, these hallucinated Metasploit paths may point to non-existent modules or to modules with incorrect semantics (wrong module type, wrong target, or mismatched options). This is not a trivial error: penetration testing relies on exact identifiers and parameter names. Even small inaccuracies, such as a mistyped module path or incompatible payload, can silently break automation, waste time, or trigger unintended actions. Such errors undermine the reproducibility and safety of LLM-guided security workflows and heighten legal and accountability risks when outputs are unverified.

\begin{figure}[h]
\centering
\includegraphics[width=1\linewidth]{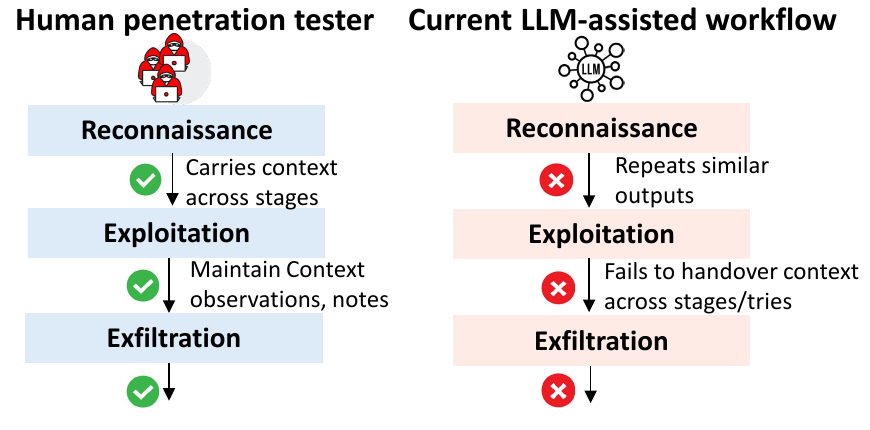}
\caption{Humans carry context across stages, while LLMs repeat outputs and fail to hand over context.}
\label{fig:memory_context}
\end{figure}

\subheading{Challenge \#2: Insufficient long-term contextual memory.}
Multi-step penetration tests require preserving accurate, versioned state across actions (reconnaissance → exploitation → exfiltration). Many LLM-assisted systems lose or overwrite prior context, repeat failed actions, or fail to incorporate intermediate findings, which reduces effectiveness and increases wasted probing. This shortfall matches evidence that attention-based models emphasize recent tokens and struggle with long-range procedural state \cite{vaswani2023attentionneed,chatGPTisNotEnough}. As shown in Figure~\ref{fig:memory_context}, human testers carry forward artifacts (discovered services, credentials, outputs) to guide later decisions, whereas LLMs often lose that thread and require frequent human intervention, limiting fully automated red-teaming.

To address these challenges, APT-Agent introduces two core contributions: a \textbf{rectification module} to mitigate Challenge \#1: hallucination of technical entities and a command-specific, stage-aware \textbf{context management module (CMM)} to overcome Challenge \#2: insufficient long-term contextual memory. (1) The rectification module validates LLM-generated commands (mainly Metasploit modules) against a database of legitimate modules and commands. It employs a hybrid correction method, first applying a fuzzy string matching algorithm (RapidFuzz ratio-based similarity) \cite{fuzzy_search} to the final segment of the hallucinated path, followed by exact suffix matching. This design strikes a balance between tolerance for minor textual deviations and the precision required for executable commands, achieving substantially higher correction rates than other approaches. (2) For context awareness, APT-Agent employs a stage-aware CMM that persistently tracks the current operational state, historical commands, and their outcomes across multiple interaction steps. Instead of passively logging outputs, it encodes prior actions and integrates history attempts into subsequent prompts. This enables adaptive strategy refinement and prevents repeated execution of invalid commands, thereby mitigating the long-horizon accuracy problem in multi-step penetration testing. Together, these modules enable APT-Agent to achieve reliable, fully automated execution across multiple pen-testing stages. We evaluate APT-Agent on Metasploitable 2 \cite{Metasploitable2}, targeting seven vulnerable services spanning web, database, and network protocols. The framework achieves an 84.29\% end-to-end exploitation success rate, substantially exceeding the performance range reported in prior LLM-based penetration testing systems \cite{b6,b7,b5}.
The source code for APT-Agent is available in our \href{https://github.com/Guanting-Li/APT-Agent_single_host.git}{GitHub repository}.

The main contributions of this work are summarized as follows:

\begin{itemize}
\item A fully automated LLM-driven red-teaming framework executing reconnaissance, exploitation, and post-exploitation with minimal human intervention.
\item A hybrid rectification module leveraging targeted fuzzy mechanisms to recover hallucinated module names and commands.
\item A stage-aware CMM that preserves state and failure history for accurate multi-step planning.
\item A comprehensive evaluation on Metasploitable-2 showing >40\% relative improvement in success rate across seven services compared to prior LLM approach.
\end{itemize}

The rest of this paper is organized as follows: Section~\ref{back_related} introduces background and related work. Section~\ref{proposed} presents our proposed work. Section~\ref{eval} presents our evaluation results. Section~\ref{limit_and_future} introduces limitations and future work related to our proposed work. Finally, Section~\ref{conclusion} concludes this paper.

\section{Background and Related Work}
\label{back_related}
This section provides an overview of the penetration testing, target environment, followed by prior research examining automated pen-testing solutions. 

\begin{table*}[t]
\centering
\resizebox{\textwidth}{!}{%
\begin{tabular}{l l c c c c}
  \toprule
  \textbf{Category} & \textbf{Work} & \textbf{Handles Hallucination} &
  \textbf{Long-Horizon Memory} & \textbf{Full Automation} & \textbf{Low Expert Setup} \\
  \midrule
  \multirow{4}{*}{\textbf{RL}}
  & Schwartz et al.\ (2020) \cite{b13} & \na & \bad & \good & \bad \\
  & Chen et al.\ (2023) GAIL-PT \cite{b14} & \na & \bad & \good & \bad \\
  & Becker et al.\ (2024) \cite{b15} & \na & \bad & \good & \bad \\
  & Li et al.\ (2023) \cite{b16} & \na & \bad & \good & \bad \\
  \midrule
  \multirow{5}{*}{\textbf{LLM}}
  & PentestGPT (2023) \cite{b6} & \bad & \bad & \bad & \bad \\
  & Script Kiddie (2023) \cite{b5} & \bad & \bad & \bad & \good \\
  & AutoAttacker (2024) \cite{b7} & \bad & \bad & \bad & \good \\
  & PentestAgent (2024) \cite{shen2025pentestagentincorporatingllmagents} & \bad & \bad & \bad & \bad \\
  & \textbf{APT-Agent (ours)} & \good & \good & \good & \good \\
  \bottomrule
\end{tabular}%
}
\vspace{0.2em}
\caption{Comparison of RL- and LLM-based automated penetration testing. “Handles Hallucination” maps to \emph{Challenge \#1}; “Long-Horizon Memory” maps to \emph{Challenge \#2}. \good\ = present/supported, \bad\ = absent/limited, \na\ = not applicable. “Low Expert Setup” indicates minimal reliance on domain experts or preconfigured environments.}
\label{tab:rl-llm-comparison}
\vspace{-0.8em}
\end{table*}

\subsection{Key Phases of Penetration Testing}

APT-Agent focuses on three primary phases of penetration testing: (1) reconnaissance, where the tester collects information about hosts, services, and software versions to construct an attack surface map; (2) exploitation, where identified vulnerabilities are actively leveraged to gain unauthorized access; and (3) post-exploitation and exfiltration, where sensitive data may be accessed, persisted, or extracted.

These phases are operationalized through frameworks such as \textit{Metasploit} \cite{Metasploit}, which provides a comprehensive library of reconnaissance, exploitation, and post-exploitation modules~\cite{b9}. Metasploit’s modular architecture offers comprehensive options for scanning, exploit execution, payload deployment, and post-exploitation activities, including credential dumping and lateral movement. This versatility makes it particularly suitable for evaluating the effectiveness of automated penetration testing approaches \cite{b6,b7}.

\subsection{Existing Automated Penetration Testing Solutions}

Research on automated penetration testing \cite{AutoPentestingAnOverview} has primarily followed two directions: reinforcement learning (RL)-based methods and LLM-based methods.

\subheading{Reinforcement Learning (RL) Approaches}. RL-based frameworks typically formalize penetration testing as a sequential decision-making problem within a Markov Decision Process (MDP), where agents learn optimal attack paths through interaction and reward feedback. Schwartz et al. \cite{b13} extended this paradigm using a partially observable MDP (POMDP) \cite{SarrauteCarlos2013PT=P} with an information decay factor, modeling how defender countermeasures gradually reduce attacker knowledge over time. This formulation enables planning under uncertainty but was validated only in simulated settings. Chen et al. \cite{b14} proposed GAIL-PT, which combines expert demonstrations with generative adversarial imitation learning to accelerate attack planning, while Becker et al. \cite{b15} benchmarked multiple RL algorithms (Q-learning, DQN, A3C) on the NASim simulator for systematic comparison. Li et al. \cite{b16} introduced a hierarchical deep reinforcement learning framework that incorporates expert prior knowledge into state and action representations, improving learning efficiency and structured planning. Despite these advances, \textit{RL-based approaches remain heavily reliant on simulation fidelity and struggle to generalise to realistic penetration testing environments.}

\subheading{LLM-Based Approaches}. LLM-driven frameworks leverage the generative and contextual reasoning capabilities of models such as GPT-3.5 \cite{GPT-3.5} and GPT-4o \cite{GPT-4o}. PentestGPT \cite{b6} introduced the Pentesting Task Tree (PTT), which modularises reasoning, generation, and parsing to decompose penetration testing into subtasks. AutoAttacker \cite{b7} expanded this paradigm with a Planner, Navigator, and Summarizer pipeline, extending automation into post-exploitation actions such as persistence and privilege escalation. Script Kiddie \cite{b5} employed a zero-shot prompt chaining strategy for end-to-end penetration testing, while PentestAgent \cite{shen2025pentestagentincorporatingllmagents} adopted multi-agent collaboration and retrieval-augmented generation (RAG) to enhance exploitation planning. These works demonstrate the adaptability of LLMs across reconnaissance, exploitation, and post-exploitation phases. However, \textit{they remain constrained by persistent issues such as hallucinated commands, limited long-term memory, and reliance on pre-configured environments or human corrections.}

\subheading{Limitations and Research Gap.}
Table~\ref{tab:rl-llm-comparison} contrasts prior RL- and LLM-based systems along four criteria aligned with our challenges: \emph{Handles Hallucination} (Challenge~\#1), \emph{Long-Horizon Memory} (Challenge~\#2), \emph{Full Automation}, and \emph{Low Expert Setup}. 

RL approaches (e.g., \cite{b13,b14,b15,b16}) score well on automation within simulators, yet they require substantial expert engineering (environment design, reward shaping), leading to limited real-world deployability and a poor ''low-expert'' profile. They do not have the hallucination problem since free-form command generation does not occur in these frameworks. 

LLM systems (e.g., \cite{b6,b5,b7,shen2025pentestagentincorporatingllmagents}) operate on real targets with lighter setup, but most lack explicit mechanisms for hallucination correction and long-horizon memory. As a result, they often require human-in-the-loop steering and fail to achieve dependable end-to-end automation.

No prior work simultaneously satisfies all four desiderata. This gap motivates our design of \textit{APT-Agent}, which retains the adaptability of LLMs while adding an explicit rectification module to address Challenge~\#1 and a structured CMM for Challenge~\#2, thereby enabling fully automated campaigns with minimal expert setup.


\begin{figure*}[t]
    \centering
    \includegraphics[width=\textwidth]{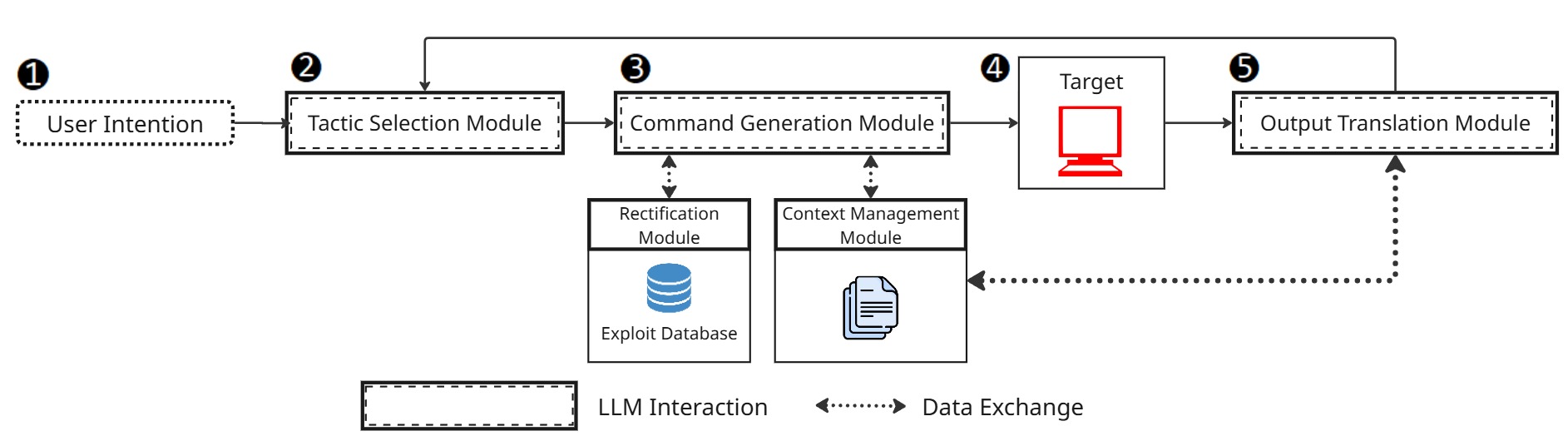}
    \caption{Overview of APT-Agent}
    \label{fig:art_agent_overview}
\end{figure*}

\section{Proposed Work}
\label{proposed}

This section presents the design overview of the overall architecture of APT-Agent, including its core modules and supporting mechanisms.
\subsection{System Overview} 
To address the limitations identified earlier, we propose \textbf{APT-Agent}, a fully automated pen-testing framework that leverages specialized LLM-driven chains to conduct penetration testing in a structured and adaptive manner. As shown in Fig.~\ref{fig:art_agent_overview}, APT-Agent consists of three core components: the \textit{Tactic Selection Module}, which determines the next phase of the MITRE ATT\&CK framework~\cite{mitre_attack}; the \textit{Command Generation Module}, which produces executable system commands or Metasploit operations; and the \textit{Output Translation Module}, which condenses tool outputs into structured insights for subsequent reasoning. These modules operate iteratively, with each cycle representing a single penetration testing step and considered as an iteration. In its current form, APT-Agent operates across three stages (\textbf{RECON}, \textbf{EXPLOIT}, and \textbf{EXFILTRATE}).

APT-Agent is designed to overcome two major weaknesses of prior LLM-based frameworks: (i) \textbf{hallucination of technical entities}, where models generate non-existent commands or module names, and (ii) \textbf{loss of contextual memory} across multi-step campaigns. To address these, APT-Agent incorporates two additional modules: a \textbf{Rectifier Module}, which validates and corrects hallucinated commands against a curated Metasploit database using hybrid fuzzy matching; and a \textbf{CMM}, which maintains stage-specific logs of executed commands and outcomes, enabling long-term context retention and adaptive reasoning.

Mirroring real-world penetration testing teams, APT-Agent separates strategic reasoning from tactical execution. The Tactic Selection Module acts as the controller, the Command Generation Module issues commands, and the Output Translation Module structures outputs before reinjecting them into the reasoning loop. This modularity ensures both specialization and coherence across the testing process.
Next, we detail each of APT-Agent’s core components, describing their design and role in enabling reliable, automated penetration testing.

\subsection{Modules in APT-Agent}
APT-Agent operates through a continuous loop of different modules: 1. Tactic Selection Module, 2. Command Generation Module and 3. Output Translation Module.

(1) The Tactic Selection Module governs the overall progression of the campaign by selecting the most appropriate stage of the MITRE ATT\&CK tactics based on accumulated knowledge and prior outcomes. The logic first checks if the campaign objective has been met. If so, the operation terminates. Otherwise, the module selects one of the stages based on the campaign's current progress. (2) The Command Generation Module transforms high-level tactic decisions into executable commands tailored to the current stage. During reconnaissance, it generates scanning or enumeration commands to reveal target information. In the exploitation stage, the LLM selects a vulnerable service based on reconnaissance results, outputs relevant details (IP, port, service, version), and produces appropriate Metasploit commands. Crucially, the module interacts with the CMM to avoid repeating failed exploits and adapt strategies dynamically. In the exfiltration stage, it adapts commands to the active session type: generating Unix commands for shell sessions and Metasploit-native commands for meterpreter sessions. This design ensures flexibility across different access conditions while maintaining progress toward the ultimate goal of sensitive file exfiltration. (3) The Output Translation Module serves as an interpretive layer that condenses raw tool outputs into structured, goal-oriented feedback. Direct exposure to verbose or noisy outputs risks misleading the LLM and propagating hallucinations. Instead, this module applies a labeling scheme to classify outcomes as \texttt{SUCCESS} or \texttt{FAIL}, summarizes the key findings, and suggests the next logical step. 

\subsection{Rectification Module}

A key innovation of APT-Agent is the \textbf{Rectification Module}, designed to mitigate a recurring weakness of LLMs: hallucinated LLM outputs. During experiments, we observed that LLMs often produced outputs that appeared syntactically valid but were absent in the Metasploit framework. For instance, when tasked with performing SSH user enumeration, an LLM generated the invalid module: \texttt{exploit/linux/ssh/openssh\_user\_enum}
\noindent instead of the valid module:
\texttt{auxiliary/scanner/ssh/ssh\_enumusers}. Such errors disrupt command execution and propagate failures through multi-step attack chains.

To address this, APT-Agent employs a \textbf{rectification module} comprising two complementary components, each detailed in the subsections that follow. The first is a curated \emph{knowledge base} of valid Metasploit modules, which serves as the authoritative reference set against which every LLM-generated path is checked. The second is the \emph{Hybrid Rectification Method}, an algorithm that maps each hallucinated module path back to an entry in the knowledge base by combining approximate fuzzy matching with suffix-based correction. Together, these components ensure that linguistically plausible but invalid LLM outputs are grounded in verifiable, executable modules.

Beyond correcting hallucinated module paths, the rectification module also performs \emph{execution-level normalization} prior to runtime. This includes injecting missing mandatory options (e.g., \texttt{RPORT}, \texttt{LHOST}, \texttt{LPORT}) and validating payload architecture against the selected module target. These checks ensure that rectified commands are not only syntactically valid but also executable within the target context.

\subsubsection{Knowledge Base Foundation}
At its core, the rectifier relies on a database that we curated and constructed of 3,253 valid Metasploit modules, indexed by attributes such as service, type, operating system, rank, and description. This structured knowledge base provides the authoritative reference for all rectified outputs.

\subsubsection{Hybrid Rectification Method}
The Hybrid Method integrates two complementary ideas: fuzzy search and suffix-level alignment, into a unified correction pipeline shown in Algorithm~\ref{alg:hybrid}:

\subheading{(a) Suffix Extraction:} The generated path $p$ is decomposed by ``/'' and the last component $s$ (e.g., \texttt{openssh\_user\_enum}) is extracted. This captures the LLM’s intended function, which is typically preserved in the suffix even when the hierarchy is incorrect.

\subheading{(b) Suffix Fuzzy Matching:} The rectification module performs fuzzy matching between $s$ and all module suffixes in the database $\mathcal{D}$ using normalized Levenshtein similarity:
    \[
    \text{sim}(x,y) = 1 - \frac{d_{\text{lev}}(x,y)}{\max(|x|,|y|)}.
    \]
    The most similar suffix $s_f$ is selected.
    
\subheading{(c) Module Reconstruction:} Once $s_f$ is identified, the rectification module retrieves the full module path in $\mathcal{D}$ that contains $s_f$ and replaces the LLM’s hallucinated output with this verified path.

\begin{algorithm}[!t]
\caption{Hybrid Rectification}
\label{alg:hybrid}
\small
\begin{algorithmic}[1]
\Require LLM path $p$, DB $\mathcal{D}$; threshold $\tau_h$
\Ensure $m^*$ or \texttt{NO\_MATCH}
\State $s \gets$ last segment of $p$
\State $s_f \gets \arg\max_{t \in \mathrm{Suffixes}(\mathcal{D})} \mathrm{sim}(s,t)$
\State \Return module in $\mathcal{D}$ with suffix $s_f$
\end{algorithmic}
\end{algorithm}


This hybrid design unifies the strengths of fuzzy and suffix-based rectification: fuzzy similarity provides tolerance to minor textual variations, while suffix grounding prevents errors from hierarchical hallucinations. Empirically, this approach achieved the highest rectification success rate among all variants, as detailed in Section~\ref{component evaluation}.

\subsection{Context Management Module (CMM)}

Multi-step penetration testing requires maintaining reliable context across stages, yet LLMs are prone to short-term memory loss and duplicated LLM outputs. Without explicit context management, prior failures or successes are easily forgotten, leading to repeated commands and inefficient exploration. While LangChain offers a composable framework for chaining LLM prompts and tools, its default workflows are stateless unless explicit memory components are introduced. Naïve approaches, such as storing entire transcripts, proved noisy and costly in token usage. Through experimentation, we found that a \textit{minimalist, high-signal log format} yields better performance by capturing only what is essential for decision-making. Hence, the CMM adopts this principle, focusing on concise, high-value state tracking for efficient decision-making. The remainder of this subsection details how this principle is realized: first, the design choices that shape the log format with the formal memory schema; second, illustrative JSON examples that demonstrate the resulting representation in practice; and third, the stage-aware router that injects this context back into the Command Generation Module at each iteration.

\subsubsection{Design Choices}
The design of the CMM was shaped by two practical considerations:

\subheading{(1) Deterministic RECON stage.} We observe that reconnaissance consistently succeeds in detecting target services and versions in our evaluation. Since these results are stable and reproducible, maintaining a RECON log would add overhead without contributing new information. 

\subheading{(2) Signal-to-noise and cost trade-off.} To avoid verbose logs that increase token consumption, the CMM records only three fields per iteration: the stage iteration number, the issued command, and its binary outcome (\texttt{success} or \texttt{fail}). Crucially, raw outputs (stdout/stderr) are excluded, as they tend to overwhelm prompts with irrelevant noise, while the essential decision-making signal is captured through the binary outcome.

Formally, the global memory state $M$ is defined as a set of stage-specific JSON logs:
\[
M = \{ M_{\text{EXPLOIT}},\; M_{\text{EXFILTRATE}} \},
\]

where each stage log $M_{\text{stage}}$ is an array of compact entries:
\[
M_{\text{stage}} = [e_1, e_2, \dots, e_k],
\]

with schema
\[
e=\left\{
\begin{array}{@{}l@{\ :\ }l@{}}
\texttt{"iter"}   & i, \\
\texttt{"cmd"}    & \text{string}, \\
\texttt{"result"} & \{\texttt{"success"},\ \texttt{"fail"}\}
\end{array}
\right.
\]
\subsubsection{Illustrative JSON Examples}
To demonstrate the compactness of this representation, we provide example log fragments below. Each entry captures only the iteration index, execution attempt and its binary outcome, discarding extraneous text:

\begin{lstlisting}[breaklines=true, basicstyle=\ttfamily\footnotesize]
M_EXPLOIT = [
  {"iter": 1, "cmd": "exploit/multi/ssh/sshexec",
   "result": "fail"},
  {"iter": 2, "cmd": "exploit/unix/ftp/vsftpd_234_backdoor",
   "result": "success"}
]
\end{lstlisting}

\begin{lstlisting}[breaklines=true, basicstyle=\ttfamily\footnotesize]
M_EXFILTRATE = [
  {"iter": 1, "cmd": "search -f flag.txt",
   "result": "success"},
  {"iter": 2, "cmd": "download /home/msfadmin/flag.txt",
   "result": "fail"}
]
\end{lstlisting}

\subsubsection{Stage-Aware Routing and Prompt Injection}
A lightweight router manages read/write access to these logs. Unlike summarization approaches that feed condensed notes back into prompts, APT-Agent directly injects the JSON log of the current stage into the \textbf{Command Generation Module}. This is feasible because the log entries are deliberately compact and easy for the LLM to parse.

This strategy has two main advantages:
\begin{itemize}
    \item \textbf{Prevention of redundancy.} The explicit \texttt{result} field ensures that previously failed commands are not re-issued.
    \item \textbf{Preservation of chronology.} Iteration numbers and raw command strings maintain execution order, giving the LLM precise historical grounding.
\end{itemize}

By balancing compactness with fidelity, the CMM enables long-horizon campaigns to retain the essential context needed for adaptive decision-making, without inflating costs or introducing noise.

\subsection{Full-Automation Mechanisms}

\textit{APT-Agent} enables fully autonomous operations by integrating mechanisms for tool execution, module configuration, runtime control, and session management, allowing seamless progression through all stages in the campaign.

\subheading{(1) Tool adapters for executable actions.}
We designed lightweight adapters that enable seamless integration with scanning and exploitation tools. 
\emph{Nmap adapter} parses multi-line model outputs, extracts the first valid 
\texttt{nmap}/\texttt{sudo nmap}/\texttt{ping} line, tokenizes it with \texttt{shlex}, enforces a timeout, and returns the execution result.
For exploitation, we rely on MSFRPCD, Metasploit’s RPC daemon, which exposes console, module, session, and job control via a programmatic API. 
The \emph{Metasploit adapter} leverages this interface to execute actionable commands and stream only relevant lines to the RPC console, avoiding verbose noise.

\subheading{(2) Metasploit Module Rectification and Setup.}
During the \textit{EXPLOIT} stage, the model outputs a header (IP/service/version/port) and a candidate \texttt{use <module>}. 
We canonicalize the service, query a local \texttt{modules} table, and perform the rectification.
Module \emph{options and payloads} are then fetched live via MSFRPCD and injected into an LLM prompt, which produces an executable, placeholder-free block (including payloads, required options, and potential default/blank credentials or a supplied wordlist). 
This eliminates the need for human operators to manually search for modules or craft option strings.

\subheading{(3) Brute-force awareness and time budgeting.}
The option-setup process automatically tags modules as brute-force or non–brute-force. This classification enables dynamic adjustment of console execution windows (e.g., 180\,s for brute-force modules vs.\ 30\,s otherwise) and tail-trimmed transcripts to manage extensive outputs, while campaign termination is governed by a fixed iteration budget. As a result, idle waiting is reduced, and noisy inputs are minimized.

\subheading{(4) Autonomous session handling.}
Beyond console-based checks, \textit{APT-Agent} actively monitors the RPC session inventory (\texttt{client.\allowbreak sessions.\allowbreak list}) to detect new sessions spawned during exploitation. 
When the foothold is a raw shell, it invokes \texttt{sessions -u <id>} to attempt upgrading into a meterpreter session—Metasploit’s in-memory post-exploitation agent with advanced file, process, and network control.\\
By combining tool-specific adapters, robust module rectification and setup, brute-force–aware option synthesis, and autonomous session management, \textsc{APT-Agent} achieves full end-to-end operation with \emph{no human intervention} beyond specifying the target IP.


\begin{table*}[ht]
\centering
\scalebox{0.9}{ 
\renewcommand{\arraystretch}{1.2}
\setlength{\tabcolsep}{6pt}
\begin{tabular*}{\textwidth}{@{\extracolsep{\fill}} l c c c c c}
\toprule
\makecell[c]{\textbf{Service}} &
\makecell[c]{\textbf{Success}} &
\makecell[c]{\textbf{Avg. Total}\\\textbf{Iterations}} &
\multicolumn{3}{c}{\textbf{Avg. Iterations per Stage}} \\
\cmidrule(lr){4-6}
 & & & \textbf{RECON} & \textbf{EXPLOIT} & \textbf{EXFILTRATE} \\
\midrule
vsftpd 2.3.4   & 10 & 5.1  & 1.5 & 1.2 & 2.4 \\
OpenSSH 4.7    & 9 & 7.6 & 1.4 & 4.2 & 2.0  \\
Telnet         & 8 & 4.6 & 1.1 & 1.2 & 2.3   \\
Apache 2.2.8   & 7 & 17.3 & 2.4 & 7.9 & 7.0   \\
UnrealIRCd     & 7 & 8.9  & 2.7 & 4.1 & 2.1   \\
PostgreSQL     & 9 & 8.6 & 1.0 & 1.5 & 6.1 \\
Samba          & 9 & 8.8 & 1.2 & 2.4 & 5.2  \\
\bottomrule
\end{tabular*}
} 
\vspace{0.4em}
\caption{Service-wise number of successes out of 10 trials, average iteration counts per stage.}
\label{tab:service_results}
\end{table*}

\section{Evaluation}
\label{eval}

We evaluate APT-Agent in a controlled Metasploitable~2 environment. Our evaluation presents service-wise performance results, compares APT-Agent with prior systems, and conducts ablation studies on key components, concluding with an analysis of the rectification and memory designs.

\subsection{Target Environment}
We set up a target environment consisting of a \textit{Metasploitable~2} virtual machine that contains a sensitive file named \texttt{flag.txt}. \textit{Metasploitable 2} is a widely adopted open-source virtual machine intentionally designed with numerous security vulnerabilities for research and training purposes~\cite{b6,b7}. The virtual machine hosts numerous vulnerable services across different layers, including web applications, databases (MySQL, PostgreSQL), and network protocols (e.g., HTTP, VSFTPD). This diversity provides a realistic attack surface spanning reconnaissance, exploitation, and post-exploitation scenarios, making it particularly suitable for benchmarking the reliability of automated penetration testing approaches.

\subsection{Experimental Setup and Methodology}
We use GPT-4o as the core LLM and the Metasploitable~2 virtual machine with intentionally vulnerable services to enable realistic, reproducible, and safe testing. In contrast, GPT-3.5-Turbo and Llama3-7B consistently failed to achieve successful exploitations across our experimental runs and were therefore excluded from further evaluation.

\subheading{Target Services:} We test seven representative services: \texttt{vsftpd} (FTP), \texttt{OpenSSH} (SSH), \texttt{Telnet}, \texttt{Apache} (HTTP), \texttt{UnrealIRCd} (IRC), \texttt{PostgreSQL}, and \texttt{Samba} (SMB). These are vulnerable/misconfigured in Metasploitable~2 and exploitable via Metasploit, offering diverse vectors (RCE, weak credentials, etc.). For brute-force dependent services, we use a short wordlist containing the correct credential pair to bound runtime and isolate evaluation to the agent's reasoning and orchestration behavior, as exhaustive credential search is a property of the underlying tooling rather than of APT-Agent itself.

\subheading{Iteration Definition and Workflow:} One \emph{iteration} is a full pass through (i) \emph{Tactic Selection}, (ii) \emph{Command Generation}, and (iii) \emph{Output Translation}: select a stage, execute a command, and analyze the result. The agent iterates until successful file exfiltration or until the maximum iteration budget (30 iterations per stage) is exceeded.

\subheading{Scenario and Tactics:}
The objective is to retrieve a sensitive file from the target. Campaigns therefore proceed through \emph{RECON} $\rightarrow$ \emph{EXPLOIT} $\rightarrow$ \emph{EXFILTRATE}. While the framework supports more tactics (e.g., lateral movement, persistence), we constrain the scope for controlled evaluation. Early runs showed EXPLOIT is most prone to hallucinated module names, motivating the Rectifier’s focus there; the strategy generalizes to other stages as needed. Excellent-ranked Metasploit modules were retrieved from the database during the rectification.

\subheading{Autonomous Trials:} Runs are fully automated. For each service, we conduct \textbf{ten independent runs}, and each experiment is capped at 30 iterations per stage. A run ends on successful file exfiltration (success) or exceeding the iteration limit (failure). This setup measures success rate and efficiency (time/iterations) without human intervention. An example execution trace of a full APT-Agent campaign, including rectification and adaptive recovery, is provided in Appendix~\ref{app:example-campaign}.

\subsection{Results}

\subheading{Service-wise Performance:} 
Table~\ref{tab:service_results} reports per-service results across 10 independent runs per target. APT-Agent succeeded in all 10 runs on \texttt{vsftpd 2.3.4}, in 9 of 10 runs on \texttt{OpenSSH 4.7}, \texttt{PostgreSQL}, and \texttt{Samba}, in 8 of 10 on \texttt{Telnet}, and in 7 of 10 on both \texttt{Apache 2.2.8} and \texttt{UnrealIRCd}, yielding an aggregate end-to-end success rate of $84.29\%$ (59/70). Average total iterations per campaign ranged from $4.6$ (Telnet) to $17.3$ (Apache 2.2.8), reflecting differences in service complexity. The EXPLOIT stage averaged between $1.2$ iterations (vsftpd, 
Telnet) and $7.9$ iterations (Apache 2.2.8), while EXFILTRATE averaged between $2.0$ (OpenSSH 4.7) and $7.0$ iterations (Apache 2.2.8), depending on each service's data exposure surface. The single failure on OpenSSH 4.7 was attributable to continuous duplicate command generation during 
exfiltration. The two Telnet failures stemmed from session-retrieval errors, in which a remote access session was successfully spawned but APT-Agent failed to attach to and interact with it. Overall, APT-Agent demonstrates consistent reliability across heterogeneous services with bounded iteration counts.

\subheading{Comparison with related works:} 
To contextualize APT-Agent’s performance, we compare it against a state-of-the-art Script Kiddie framework~\cite{b5} and PentestGPT~\cite{b6}. All systems were evaluated under identical conditions using GPT-4o as the underlying LLM. As PentestGPT is human-in-the-loop, whereas APT-Agent and Script Kiddie operate fully autonomously, human involvement in PentestGPT was explicitly constrained for fairness. Permitted interactions were limited to copy–paste actions and simple natural-language prompts directly following PentestGPT’s own feedback, without introducing expert knowledge or external guidance. Runs requiring more than three human interventions were deemed unsuccessful, following prior work.

\begin{figure}
    \centering
    \includegraphics[width=0.7\linewidth]{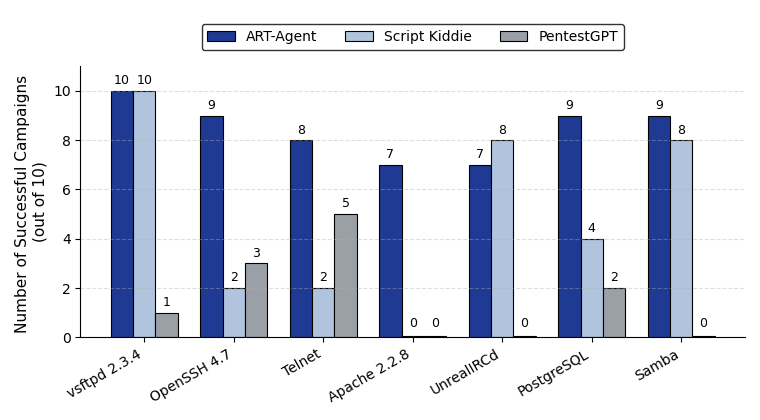}
    \caption{APT-Agent vs. Script Kiddie}
    \label{fig:art_vs_scriptkiddie_vs_pentestgpt}
\end{figure}

Fig.~\ref{fig:art_vs_scriptkiddie_vs_pentestgpt} reports service-wise results across seven target services, with ten independent runs per service. APT-Agent consistently outperforms both baselines across nearly all services. The largest gaps appear on \texttt{Apache~2.2.8}, \texttt{OpenSSH~4.7}, and \texttt{Telnet}, where PentestGPT achieves at most 0–5 successes and Script Kiddie 0–2, while APT-Agent succeeds in 7–9 runs. On services including \texttt{Apache~2.2.8}, \texttt{UnrealIRCd}, and \texttt{Samba}, PentestGPT fails to achieve any successful runs, reflecting its limited autonomy and reduced effectiveness on services and CVE-based vulnerabilities. Aggregated across all seven services, APT-Agent attains an overall success rate of 84.29\%, compared to 48.57\% for Script Kiddie and 18.57\% for PentestGPT, demonstrating stronger generalization and robustness in scenarios requiring iterative reasoning, error recovery, and adaptive command generation. A detailed contextual comparison with PenHeal~\cite{huang_penheal_2023} is provided in Appendix~\ref{app:penheal-comparison}.

\begin{figure}[t]
    \centering
    \begin{minipage}[t]{0.48\textwidth}
        \centering
        \includegraphics[width=\linewidth]{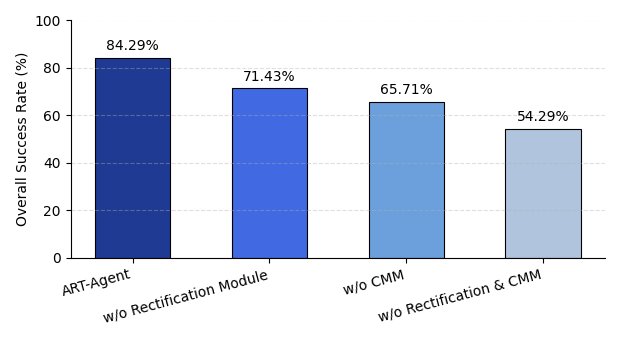}
        \caption{Ablation Study: Success Rate}
        \label{fig:ablation_success}
    \end{minipage}
    \hfill
    \begin{minipage}[t]{0.48\textwidth}
        \centering
        \includegraphics[width=\linewidth]{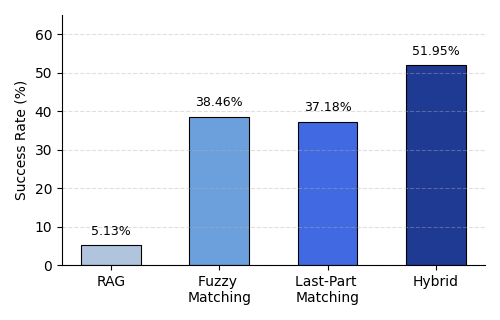}
        \caption{Rectification Methods Success Rate}
        \label{fig:Rectification Methods Success Rate}
    \end{minipage}
\end{figure}
\vspace{-1em}

\subsection{Ablation Study}

To quantify the contribution of each core component, we conducted an ablation study in which the Rectification Module and the CMM were removed individually and jointly. As shown in Fig.~\ref{fig:ablation_success}, removing either component substantially degraded overall performance. In its full configuration, APT-Agent achieved an $84.29\%$ end-to-end success rate (59/70), while omitting the Rectification Module reduced this to $71.43\%$ (50/70), and removing the CMM decreased it further to $65.71\%$ (46/70). When both modules were disabled, the success rate dropped to $54.29\%$ (38/70). Across the ablation runs, APT-Agent issued 53 Metasploit module invocations, 16 of which ($30.2\%$) were hallucinated; the Rectification Module recovered 10 of these, yielding a $62.5\%$ in-run correction rate. The CMM's contribution to efficiency was equally pronounced: on \texttt{UnrealIRCd}, average EXFILTRATE iterations fell from 52 without the CMM to 3 with it, a 17-fold reduction. Together, these results underscore the modules' complementary roles: the Rectifier recovers otherwise-failed exploit attempts, while the CMM eliminates redundant actions and stabilizes multi-step execution.

\subsection{Component-Level Evaluation}
\label{component evaluation}

We assessed various rectification methods (for hallucinated LLM outputs) and context-awareness methods (for long-term context awareness), which motivate the hybrid rectifier and the CMM.

\subheading{Rectification Methods.} 
We evaluated four rectification strategies: RAG-based retrieval, fuzzy matching,
last-part matching, and a hybrid approach. RAG struggled with structured identifiers,
while fuzzy and suffix matching failed when large portions of the path were corrupted.
The hybrid approach combines lexical tolerance with structural grounding, yielding the
highest correction rate.

Figure~\ref{fig:Rectification Methods Success Rate} compares the effectiveness of four rectification methods for recovering hallucinated module names. Each method was evaluated on 78 LLM-generated hallucinated modules to assess whether it could correct them to executable modules that preserved the original intent. The hybrid approach achieved the highest success rate (51.95\%), outperforming fuzzy matching (38.46\%), last-part matching (37.18\%), and RAG (5.13\%). The 51.95\% reflects the rectifier's intrinsic accuracy on a held-out set of 78 hallucinated module names, and is distinct from the 62.5\% in-run correction rate reported in Section V-A, which is measured over the smaller and differently-distributed set of hallucinations actually produced during live campaigns. These results demonstrate that the Hybrid method combining lexical similarity with structural pattern matching yields a more robust correction mechanism adopted as the default rectification method in APT-Agent.


\subheading{Context Awareness Evaluation.}

\subheading{(1) Conversation Buffer Memory (CBM):}
CBM stores the conversational history between the LLM and the user. We tested two variants: (i) full conversation history and (ii) results from the Output Translation Module only. Duplication was lower with full history (40.48\%) than with results-only storage (49.15\%), while both achieved 2 success runs out of 5.

\subheading{(2) Conversation Summary Buffer Memory (CSBM):}
CSBM periodically condenses history into summaries for token efficiency. In practice, this approach performed poorly, yielding 1 success run and the highest duplication rate (68.75\%). This limitation arises because summarization often omits essential details, and repeated condensation further increases the token cost per campaign.

\subheading{(3) Context Management Module (CMM):}
CMM maintains structured, stage-specific JSON logs of prior actions, explicitly recording historical commands and their outcomes, which are reinjected into subsequent prompts. This design prevents the regeneration of ineffective modules and achieved the best performance, with 5/5 successful runs and the lowest duplication rate (16.67\%).

\noindent
Given these results, \textit{APT-Agent} adopts the CMM as its default memory mechanism, enhanced with a stage-aware router that reinjects only context relevant to the active campaign phase, preserving efficiency while ensuring continuity.



\begin{figure}[t]
    \centering
    \begin{minipage}[t]{0.48\textwidth}
        \centering
        \includegraphics[width=\linewidth]{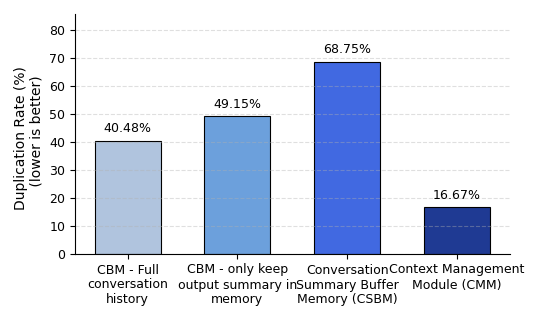}
        \caption{LLM Context Comparison: Duplication Rate}
        \label{fig:Memory_comaprison_duplication_rate}
    \end{minipage}
    \hfill
    \begin{minipage}[t]{0.48\textwidth}
        \centering
        \includegraphics[width=\linewidth]{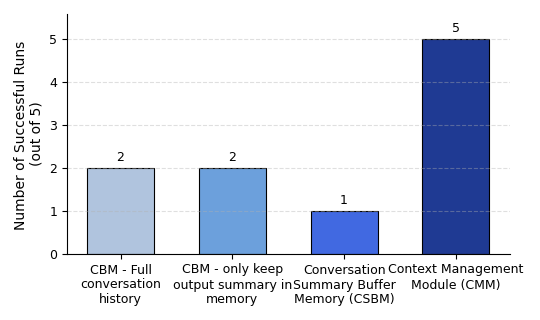}
        \caption{LLM Context Methods Comparison: Success Rate}
        \label{fig:Memory_Methods_Comparision_success_rate}
    \end{minipage}
\end{figure}

\section{Limitations and Future Work}
\label{limit_and_future}
\subheading{Role of Rectification and CMM.}
The results underscore the necessity of domain-grounded safeguards in LLM-driven security automation. The Rectification Module mitigates hallucinations—particularly during the EXPLOIT stage, while the CMM minimizes redundant commands by reinjecting stage-relevant command history. Ablation results confirm that removing either safeguard sharply reduces success rates and increases iteration counts. These findings suggest that reliability in LLM-based cyber agents depends less on model scale and more on domain-grounded architectural constraints. By coupling language reasoning with verifiable knowledge bases and explicit state tracking, APT-Agent bridges the gap between general LLM fluency and the precision required in security tooling.

\subheading{Extensibility of tactics.} 
Our experiments demonstrate that \textit{APT-Agent} can autonomously execute penetration campaigns with strong reliability in the target environment. Although the evaluation focused on RECON, EXPLOIT, and EXFILTRATE stages, the Tactic Selection Module and supporting components are readily extensible. Additional tactics defined in ATT\&CK, such as lateral movement and privilege escalation, can be incorporated by extending decision rules and augmenting command/action templates, thereby enabling broader coverage of the attack lifecycle.

\subheading{Toward more complex environments.}
Beyond expanding its tactical repertoire, future work will deploy \textit{APT-Agent} in larger and more heterogeneous environments, such as multi-host networks with interacting services and cross-machine attack paths. We also plan to evaluate its adaptability to diverse vulnerability classes (e.g., web vulnerabilities, capture-the-flag) and operational contexts, moving closer to a fully general-purpose framework for autonomous, LLM-driven pen-testing.

\subheading{Scalability and Adaptation.}
APT-Agent’s modular design allows expansion beyond its evaluated stages (RECON, EXPLOIT, EXFILTRATE). The same tactic-selection and memory logic can be extended to additional ATT\&CK phases such as lateral movement or persistence. Future experiments will deploy the framework in multi-host networks and CTF-style challenge environments to test adaptability across web, database, cloud, and kernel-level vulnerabilities, progressing toward a general-purpose autonomous red-teaming system.

\subheading{Ethical and safety considerations.} 
As highlighted in recent works~\cite{mayoralvilches2023exploitflowcybersecurityexploitation,zhang2023doesllmgeneratesecurity,abuadbba2026promise}, both defenders and adversaries are increasingly leveraging LLMs in cybersecurity operations. This dual-use nature underscores the need to strengthen defensive preparedness. Our goal in developing \textit{APT-Agent} is to uplift red-team capabilities, enabling defenders to proactively uncover vulnerabilities and harden systems before malicious actors exploit them. At the same time, responsible deployment remains essential: guardrails such as sandboxed environments, access controls, and human oversight are required to prevent misuse. Establishing standardized benchmarks, safety guidelines for automated pen-testing, and responsible disclosure practices will be critical for safe and effective adoption.

\section{Conclusion}
\label{conclusion}

In this paper, we have introduced APT-Agent, which is a fully autonomous red-teaming framework powered by LLMs and reinforced with rectification and context management modules. APT-Agent demonstrated the ability to autonomously complete penetration testing campaigns, including reconnaissance, exploitation, and exfiltration—without human intervention. Compared to prior LLM-based approaches, it achieved higher success rates, improved reliability, and reduced hallucination-induced errors by grounding outputs in executable system knowledge and maintaining stage-specific memory.  Our evaluations highlighted the scientific importance of the rectification and context management modules. The proposed rectifier mitigated invalid module generations while the memory module prevented redundant actions and enabled adaptive reasoning across campaign stages. Ablation studies confirmed that disabling either component led to substantial performance degradation, underscoring their necessity for long-horizon automated penetration testing. 

\bibliographystyle{IEEEtran}
\bibliography{bibfile}

\newpage
\appendices
\section{Comparison with PenHeal}
\label{app:penheal-comparison}
PenHeal~\cite{huang_penheal_2024} also evaluates its framework on the Metasploitable II environment, enabling a direct contextual comparison of experimental settings. However, the two systems optimize for fundamentally different objectives. PenHeal is designed to maximize vulnerability discovery coverage and to generate cost-aware remediation recommendations. In contrast, ART-Agent is exploitation-driven, prioritizing end-to-end campaign completion, including successful exploitation, error recovery, and adaptive command generation under realistic execution constraints. Consequently, ART-Agent may intentionally terminate after achieving its attack objective, rather than exhaustively enumerating all vulnerabilities. These differences reflect a deliberate trade-off between coverage-oriented security assessment and goal-driven autonomous red teaming, rather than a direct performance comparison.

\section{Example APT-Agent Campaign Log (Abbreviated)}
\label{app:example-campaign}

\begin{tcolorbox}[
  colback=gray!5,
  colframe=black,
  title=\textbf{Illustrative End-to-End APT-Agent Campaign (Abbreviated)},
  fonttitle=\bfseries,
  breakable
]
\textbf{Target:} \texttt{192.168.102.197}

\medskip
\textbf{[Iter. 1 | RECON]}\\
\texttt{nmap -sS -sV 192.168.102.197}\\
\textbf{Result:} Apache httpd 2.2.8 detected on port 80.

\medskip
\textbf{[Iter. 2 | EXPLOIT | Rectification]}\\
\textbf{Module selected:} \\\texttt{exploit/multi/http/\\apache\_mod\_cgi\_bash\_env\_exec}

\textbf{Raw command:}
\begin{lstlisting}
use exploit/multi/http/apache_mod_cgi_bash_env_exec
set RHOSTS 192.168.102.197
set PAYLOAD linux/x86/meterpreter/reverse_tcp
exploit
\end{lstlisting}

\textbf{Rectifier actions:}
(i) injects missing required option (\texttt{RPORT});\\
(ii) validates payload architecture against the module target.

\textbf{Rectified command:}
\begin{lstlisting}
use exploit/multi/http/apache_mod_cgi_bash_env_exec
set RHOSTS 192.168.102.197
set RPORT 80
set PAYLOAD linux/x86/meterpreter/reverse_tcp
exploit
\end{lstlisting}

\textbf{Outcome:} Execution failed; module recorded.

\medskip
\textbf{[Iter. 3 | EXPLOIT | Rectification \& Recovery]}\\
\textbf{Raw module reference:} \texttt{exploit/unix/webapp/\\php\_cgi\_arg\_injection}

\textbf{Rectifier actions:}
(i) resolves the LLM-generated module reference to the canonical Metasploit path;\\
(ii) injects missing mandatory options (\texttt{RPORT}, \texttt{LHOST}, \texttt{LPORT});\\
(iii) confirms payload compatibility with the PHP CGI execution context.

\textbf{Rectified executable module:}\\
\texttt{exploit/multi/http/\\php\_cgi\_arg\_injection}

\textbf{Outcome:} \textbf{Meterpreter session established} (\texttt{www-data}).

\medskip
\textbf{[Iter. 4--6 | EXFILTRATE]}
\begin{lstlisting}
search -f flag.txt
search -d / -f flag.txt
cat /home/msfadmin/flag.txt
\end{lstlisting}
\textbf{Outcome:} Sensitive file successfully retrieved.

\medskip
\textbf{[Iter. 7 | END\_OF\_CAMPAIGN]}\\
Goal achieved; campaign terminated.
\end{tcolorbox}

\begin{tcolorbox}[
  colback=gray!5,
  colframe=black,
  title=\textbf{Exploitation Memory ($M_{\mathrm{EXPLOIT}}$)},
  fonttitle=\bfseries,
  breakable
]
\begin{lstlisting}
[
  {iter: 2,
   cmd: apache_mod_cgi_bash_env_exec,
   result: fail},
  {iter: 3,
   cmd: php_cgi_arg_injection,
   result: success}
]
\end{lstlisting}
\end{tcolorbox}

\begin{tcolorbox}[
  colback=gray!5,
  colframe=black,
  title=\textbf{Exfiltration Memory ($M_{\mathrm{EXFILTRATE}}$)},
  fonttitle=\bfseries,
  breakable
]
\begin{lstlisting}
[
  {iter: 4,
   cmd: search -f flag.txt,
   result: fail},
  {iter: 5,
   cmd: search -d / -f flag.txt,
   result: success},
  {iter: 6,
   cmd: cat /home/msfadmin/flag.txt,
   result: success}
]
\end{lstlisting}
\end{tcolorbox}

\end{document}